\documentclass[journal=jctcce,manuscript=article]{achemso}
\usepackage{chemformula} % Formula subscripts using \ch{}
\usepackage[T1]{fontenc} % Use modern font encodings
\usepackage{amsmath}
\usepackage{graphicx}
\usepackage{array}
\usepackage[normalem]{ulem}
\usepackage[table-number-alignment=center]{siunitx}
\newcolumntype{M}[1]{>{\centering\arraybackslash}m{#1}}
\newcommand{\onlinecite}[1]{\hspace{-1 ex} \nocite{#1}\citenum{#1}}

\author{Piotr Michalak}
\author{Michał Lesiuk}
\affiliation[University of Warsaw]
{Faculty of Chemistry, University of Warsaw, Pasteura 1, Warsaw, 02-093, Poland}
\email{m.lesiuk@uw.edu.pl}

\title{Rank-reduced equation-of-motion coupled cluster triples: an accurate and affordable way of calculating electronic excitation energies}

\begin{document}

\begin{tocentry}
\includegraphics[width=8.0cm, height=4.4cm]{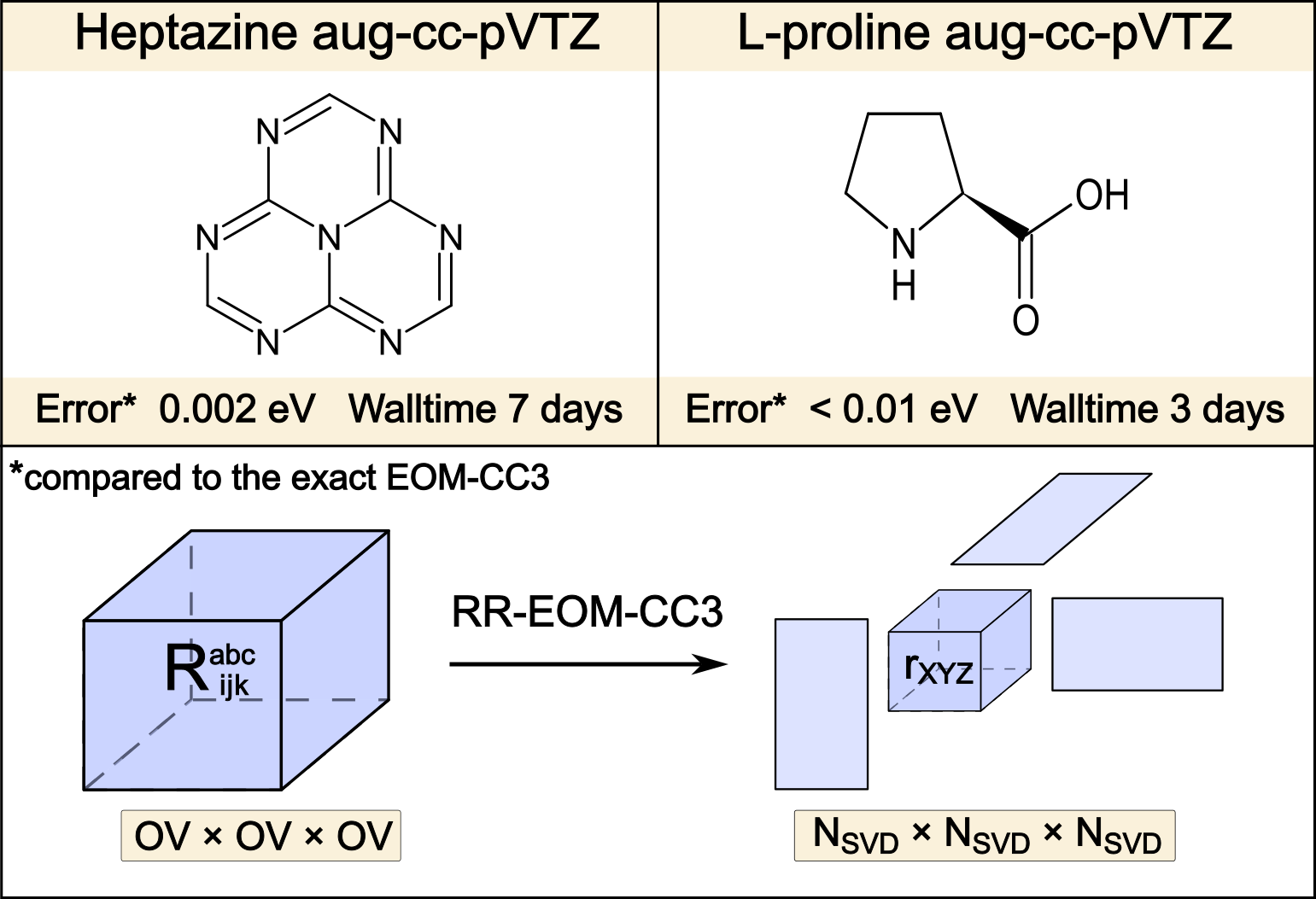}
\end{tocentry}

\begin{abstract}
 In the present work we report an implementation of the rank-reduced equation-of-motion coupled cluster method with approximate triple excitations (RR-EOM-CC3). The proposed variant relies on tensor decomposition techniques in order to alleviate the high cost of computing and manipulating the triply-excited amplitudes. In the RR-EOM-CC3 method, both ground-state and excited-state triple-excitation amplitudes are compressed according to the Tucker-3 format. This enables to factorize the working equations such that the formal scaling of the method is reduced to $N^6$, where $N$ is the system size. An additional advantage of our method is the fact the accuracy can be strictly controlled by proper choice of two parameters defining sizes of triple-excitation subspaces in the Tucker decomposition for the ground and excited states. Optimal strategies of selecting these parameters are discussed. The developed method has been tested in a series of calculations of electronic excitation energies and compared to its canonical EOM-CC3 counterpart. Errors several times smaller than the inherent error of the canonical EOM-CC3 method (in comparison to FCI) are straightforward to achieve. This conclusion holds both for valence states dominated by single excitations and for states with pronounced doubly-excited character. Taking advantage of the decreased scaling, we demonstrate substantial computational costs reductions (in comparison with the canonical EOM-CC3) in the case of two large molecules -- L-proline and heptazine. This illustrates the usefulness of the RR-EOM-CC3 method for accurate determination of excitation energies of large molecules.
\end{abstract}
\newpage

\section{Introduction}
\label{sec:intro}

Excited electronic states of polyatomic molecules play a crucial role in many branches of science such as photochemistry~\cite{garavelli2006,mai2020}, optical spectroscopy~\cite{grimme2004,barone2007,perkampus2013} or atmospheric physics~\cite{vereecken2012}. In general, as a molecule is excited into a higher electronic state, its properties may change dramatically due to the excess energy available. As a result, many chemical and physical processes prohibited in the ground state become possible after the excitation. 

Theoretical studies of properties of excited states are crucial to complement the experimental work and to explain and rationalize the observations. Therefore, it is not surprising that development of theoretical methods capable of accurate and reliable description of molecular excited states has been the focus of many researchers~\cite{serrano2005,dreuw2005,gonzalez2012,lischka2018,mai2020}. Focusing here on single-reference methods, conceptually the simplest method of this kind is the configuration interaction singles (CIS)~\cite{bene1971,ditchfield1972,foresman1992}. Unfortunately, the accuracy offered by this method is insufficient in many situations and hence approximate corrections that take double excitations into account, such as CIS(D) or CIS(D$_\infty$), were developed~\cite{head1994,head1995,head1999}. Another popular method is the time-dependent density functional theory~\cite{casida1995,dreuw2005,casida2012} (TDDFT) which offers an attractive accuracy-to-cost ratio and has found a widespread use. However, the TDDFT results are often sensitive to the choice of the exchange-correlation functional and careful comparison with benchmark calculations or experimental data is necessary~\cite{laurent2013,bremond2018,loos2019,sarkar2021}. Next, the algebraic-diagrammatic construction (ADC) method has seen a surge in popularity in recent years~\cite{schirmer1982,dreuw2015,herbst2020,dreuw2023}. While this method is perturbative in nature, it has a number of advantages such as the use of explicitly hermitian Hamiltonian which provides a straightforward access to excited-states properties necessary in many applications. Finally, the equation-of-motion coupled cluster (EOM-CC) family offers a systematically improvable hierarchy of methods which can be truncated at a given excitation level~\cite{sekino1984,stanton1993,krylov2008,bartlett2010,sneskov2012,bartlett2012}. For an extended discussion of the aforementioned, as well as other, quantum chemistry methods targeting the molecular excited states see the recent reviews~\cite{serrano2005,dreuw2005,gonzalez2012,lischka2018,mai2020}.

This work focuses on the EOM-CC approach and, in particular, on excited-state methods that include triple excitations with respect to the reference determinant. It is accepted that triple excitations are required for benchmark-quality results and for description of states with significant double-excitation character, as an example~\cite{loos2018,loos2020,loos2021}. In the case of the ground state, the triple excitations are most commonly included by the means of the CCSD(T) method~\cite{purvis82,scuseria87,ragha89}, widely seen as the "gold standard" of quantum chemistry. The (T) correction, which is perturbative in nature, is added on top of the converged CCSD energy and such approach is much less computationally costly than the complete CCSDT model~\cite{noga87,scuseria88}. Unfortunately, in the case of the excited states, definition of an analogous (T) correction appears to be significantly more difficult~\cite{watts1995,watts1996,stanton1996,christiansen1996,hirata2001,kowalski2004,manohar2009,sauer2009,watson2013,matthews16}. An alternative to such perturbative approach is offered by the CC3 model~\cite{christiansen1995,koch1997}. In this method the equations that determine the triply-excited amplitudes are approximated (in comparison with CCSDT), but the iterative nature of the method is retained. In recent years, the CC3 method has been frequently used as \emph{de facto} benchmark-quality standard for medium-sized systems, capable of delivering accuracy of the order of 0.03~eV or so~\cite{loos2018,loos2020,loos2021,véril2021,loos2021b}. An additional advantage of non-perturbative methods such as CC3 is that they enable a consistent definition of transition moments, couplings between states, etc.~\cite{christiansen1995,christiansen1998,pawlowski2005,tucholska2014,tucholska2017,Paul2021,tucholska2022} These quantities are of huge importance for many applications, especially when non-adiabatic nuclear dynamics simulations are required to reproduce experimental results~\cite{tully2012,curchod2018,mai2018,prezhdo2021}.

Despite these advantages, the EOM-CC3 is an expensive method with computational costs proportional to $N^7$, where $N$ denotes the system size. Moreover, as the size of the system increases, the large number of floating-point operations is compounded by significant memory requirements that exceed the capabilities of most machines. To alleviate these problems, the use of pair-natural orbitals~\cite{frank2020}, multilevel methods~\cite{paul2022}, and efficient implementation techniques~\cite{Paul2021}, have been suggested in the literature. This work is concerned with application of tensor decomposition techniques~\cite{kolda09} to the excited-state CC3 calculations. Tensor decomposition methods are a relatively new addition to the quantum chemistry toolbox. While the decomposition techniques applied to electron repulsion integrals, such as Cholesky decomposition~\cite{beebe77,roeggen86,koch03,folkestad19} or density fitting~\cite{whitten73,baerends73,dunlap79,alsenoy88,vahtras93}, have been widely used for a long time, application of more general techniques to coupled cluster amplitudes is a more recent development, see Refs.~\onlinecite{kinoshita03,hino04,scuseria08,bell10,hohenstein12,parrish12,parrish13a,parrish13b,mayhall17,schutski17,hohenstein19,parrish19,hohenstein2022} for representative examples. Thus far, the focus has been primarily on ground-state calculations, but applications to the EOM-CCSD theory have also been reported~\cite{hohenstein19}.

In this paper, we propose a rank-reduced variant of the CC3 method in which the triply-excited amplitudes (both for the ground and excited state) are represented in the Tucker format~\cite{tucker66}. We show that by careful optimization of the order of tensor contractions and exploiting the compression level offered by the Tucker decomposition, the scaling of the CC3 calculations can be reduced to $N^6$. This is accompanied by a significant reduction of the memory storage requirements. Equally importantly, errors introduced by the decomposition of the amplitudes can be controlled by proper selection of the excitation subspace size. With default recommended settings, the discrepancies between the rank-reduced and canonical CC3 methods are significantly smaller than the inherent errors of the latter (in comparison to FCI) for all systems considered. This is true also for difficult excited states with doubly-excited character, for which the standard EOM-CCSD fails completely with errors of the order of several eV. Finally, we point out that while the CC3 method is the focus of the present work, the same rank-reduction techniques can be applied to other iterative CC models with approximate triple excitations, such as CCSDT-$n$ family~\cite{lee1984,urban1985,noga1987}, distinguishable cluster models~\cite{kats2013,kats2014,rishi2019} or $n$CC approximations~\cite{bartlett2006,teke2024}.

\section{Preliminaries}

\subsection{Equation-of-motion coupled cluster theory}
\label{sec:eom}

Before we turn to the description of the theory, we introduce the notation adopted in this work as summarized in Table \ref{notation}. We are concerned only with electronic singlet excited states obtained from a closed-shell reference. The reference state, $|0\rangle$, is assumed to be the canonical Hartree-Fock determinant and the orbital energies are denoted by $\epsilon_p$. In all derivations we use normal-ordered electronic Hamiltonian, $H_N = H - E_{\mathrm{HF}}$, where $E_{\mathrm{HF}}$ is the Hartree-Fock energy. We assume a customary partitioning of this operator, $H_N = F_N + V_N$, where $F_N$ and $V_N$ are the normal-ordered Fock operator and fluctuation potential, respectively. The excited-state determinants are denoted by $|_i^a\rangle$, $|_{ij}^{ab}\rangle$, $|_{ijk}^{abc}\rangle$, etc., according to the excitation level. Summation over repeating indices is implied throughout the present work unless explicitly stated otherwise.

\begin{table}[t] 
\centering
\small
    \begin{tabular}{M{2cm} p{11.5cm} M{2cm}}
          indices & \centering meaning  & range \\
          \hline
        $i,j,k,l, \ldots$  & active orbitals occupied in the reference  &  $O$ \\
        $a, b, c, d, \ldots$  & orbitals unoccupied in the reference (virtual)  &  $V$ \\
        $p, q, r, s, \ldots$  & general orbitals (occupation not specified)  &  $N$ \\
        $P, Q, \ldots$  & density fitting auxiliary basis set  &  $N_{\mathrm{aux}}$ \\
        $x, y, z, \ldots$  & compressed subspace of the triply-excited ground-state amplitudes  &  $N_{\mathrm{svd}}$ \\
        $X, Y, Z, \ldots$  & compressed subspace of the triply-excited excited-state amplitudes  &  $N_{\mathrm{SVD}}$ 
    \end{tabular}
    \caption{Details of the notation used in the text.  }
    \label{notation}
\end{table}

Equation-of-motion (EOM) coupled cluster theory is a well-known approach to the calculation of excitation energies of molecules, which builds upon the ground-state coupled cluster ansatz. The EOM wavefunction of an excited state is written as:
\begin{align}
\label{eomwfn}
\Psi = R\,e^T |0\rangle.
\end{align}
In the above equation $R$ and $T$ are the cluster operators defined by the following equations:
\begin{align}
    R &= R_0 + R_1 + R_2 + \ldots + R_M, \\
    T &= T_1 + T_2 + \ldots + T_M,
\end{align}
where $R_m$ and $T_m$ are $m$-tuple excitation operators, each depending on a set of amplitudes, $R_{i_1,i_2, \ldots, i_m}^{a_1,a_2,\ldots, a_m}$ or $T_{i_1,i_2, \ldots, i_m}^{a_1,a_2,\ldots, a_m}$, with the exception of $R_0$ which is a constant. The sum in both cases extends to $M$ -- the number of electrons in the system. The amplitudes $T_{i_1,i_2, \ldots, i_m}^{a_1,a_2,\ldots, a_m}$ are usually determined by preceding ground-state CC calculations. The EOM equations for $R_{i_1,i_2, \ldots, i_m}^{a_1,a_2,\ldots, a_m}$ are found by inserting the ansatz \ref{eomwfn} into the Schr\"odinger equation (with the electronic Hamiltonian $H_N$) and projecting the equation onto the reference determinant and set of singly-, doubly-, triply-, etc., excited determinants:
\begin{align}
\label{eomeqs}
\begin{split}
    \langle 0|[\bar{H}_N, R]|0\rangle &= \omega\, \langle 0 | R |0\rangle, \\ 
    \langle _i^a|[\bar{H}_N, R]|0\rangle &= \omega\, \langle _i^a | R |0\rangle, \\
    \langle _{ij}^{ab}|[\bar{H}_N, R]|0\rangle &= \omega\, \langle _{ij}^{ab} | R |0\rangle, \\
    \langle _{ijk}^{abc}|[\bar{H}_N, R]|0\rangle &= \omega\, \langle _{ijk}^{abc} | R |0\rangle,
\end{split}
\end{align}
and so on, where $\omega$ denotes the excitation energy and $\bar{H}_N = e^{-T}H_N e^T$. The commutators in the above expressions were introduced in order to directly determine the excitation energy (rather than the total energy of a given state) from these equations. The EOM equations are, in fact, equivalent to a single eigenvalue equation for an effective Hamiltonian matrix $\mathcal{H}$, elements of which are given by $\mathcal{H}_{mn}= \langle \mu_m | [\bar{H}_N, \mu_n]|0\rangle$. Here $\mu_n$ is a collective notation for all $n$-tuple excitation operators. In this view, the amplitudes $R_i^a$, $R_{ij}^{ab}$, etc., are elements of the eigenvector corresponding to a singly-excited, doubly-excited, etc. blocks and the excitation energy $\omega$ is the corresponding eigenvalue. Notice that the $R_0$ parameter decouples from the rest of the equations and can be set to zero.

\subsection{EOM-CC3 method}
\label{sec:eomcc3}

In this work we focus on the EOM-CC3 method in which the operators $T$ and $R$ are truncated by neglecting all excitations higher than the triples, i.e. beyond $R_3$ and $T_3$. The cluster amplitudes present in the $T$ operator are determined from ground-state CC3 calculations. In the EOM-CC3 method one adopts additional approximations in the equation that determines $R_3$, while projections onto singly- and doubly-excited determinants are treated in the same way as in EOM-CCSDT. Upon expanding the $\bar{H}_N$ operator in Eqs.~\ref{eomeqs} with Baker-Campbell-Hausdorf formula (nested commutator expansion), the EOM-CC3 equations assume the following form:
\begin{align}
    &\langle _i^a| O_1^{\mathrm{CCSD}} + [\tilde{H}_N, R_3]|0\rangle = \omega\, R_i^a, \label{singleeq}\\ 
    &\langle _{ij}^{ab}|O_2^{\mathrm{CCSD}} + [\tilde{H}_N, R_3] + [[\tilde{H}_N, T_3], R_1]|0\rangle = 
    \omega\, R_{ij}^{ab}, \label{doubleeq}  \\
    &\langle _{ijk}^{abc}|[\tilde{F}_N, R_3] + [[\tilde{H}_N, T_2], R_1] + [\tilde{H}_N, R_2]|0\rangle = 
    \omega\, R_{ijk}^{abc}, \label{tripleeq}
\end{align}
where $O_1^{\mathrm{CCSD}}$ and $O_2^{\mathrm{CCSD}}$ collectively denote terms appearing in the EOM-CCSD equations -- they can be found, for example, in Refs.~\onlinecite{stanton1993, musiał2020}. In the present work we employ the $T_1$-similarity-transformed formalism in order to reduce the length of working equations. $T_1$-transformed operators are marked with a tilde and defined, for an arbitrary operator $A$, as $\tilde{A} = e^{-T_1}A\,e^{T_1}$. The corresponding $T_1$-transformed two-electron integrals are denoted by $(pq\widetilde{|}rs)$ in the Coulomb notation.

It is known that the computational costs of the EOM-CC3 method are proportional to $N^7$, where $N$ is the system size~\cite{christiansen1995, Paul2021}. By contrast, the scaling of the EOM-CCSD method is proportional to $N^6$, showing that terms involving triply-excited amplitudes are responsible for the increased scaling. To simplify the subsequent manipulations, we recall all terms appearing in the EOM-CC3 theory which depend on $T_{ijk}^{abc}$ or $R_{ijk}^{abc}$ and analyze their costs in more detail. For brevity, we define two types of permutation operators
\begin{align}
    P_2 &= \left( 1 + P_{ij}^{ab} \right), \\
    P_3 &= \left(1 + P_{ij}^{ab} \right) \left(1 + P_{ik}^{ac} + P_{jk}^{bc}  \right),  
\end{align}
where $P_{ij}^{ab}$ exchanges the compound indices $_i^a$ and $_j^b$. Projection onto singly- and doubly-excited determinants gives the following matrix elements
\begin{align} \label{firstmatrixel}
\langle_{i}^{a}| [ \tilde{H}_N , R_{3} ] |0\rangle &= 2(jc\widetilde|kb) R_{kji}^{bca}
-(jc\widetilde|kb) R_{kji}^{cba} 
- 2(jb\widetilde|kc) R_{ijk}^{bac} + (jc\widetilde|kb) R_{kji}^{abc},  
\end{align}
\begin{align} \label{secondmatrixel}
\langle_{ij}^{ab}| [ \tilde{H}_N , R_{3} ] |0\rangle &= P_2 \Big[ \widetilde{F}_{kc}R_{kji}^{cba} - \widetilde{F}_{kc}R_{kji}^{bca} -  \underline{2(ki\widetilde|lc) R_{ljk}^{cba}} + \underline{(ki\widetilde|lc)R_{ljk}^{bca}} + \underline{(kc\widetilde|li) R_{lkj}^{cab}}  \\ \nonumber
&- \uwave{(ac\widetilde|kd) R_{kji}^{bdc}} + \uwave{2(ac\widetilde|kd) R_{kji}^{dbc}} - \uwave{(ad\widetilde|kc) R_{kij}^{dcb}}\Big],
\end{align}
\begin{align} \label{thirdmatrixel}
\langle_{ij}^{ab}| \big[[ \tilde{H}_N , T_{3} ], R_1 \big]|0\rangle = 
P_2 \Big[2 (ld\widetilde|kc)T_{kji}^{cba}R_l^d - (kd\widetilde|lc)T_{kji}^{cba}R_l^d - 2(ld\widetilde|kc)T_{kji}^{bca}R_l^d  \\   \nonumber + (kd\widetilde|lc)T_{kji}^{bca}R_l^d  - \underline{2(kd\widetilde|lc)T_{kji}^{dca}R_l^b} + \underline{(kd\widetilde|lc)T_{kji}^{cda}R_l^b}  - \underline{2(kd\widetilde|lc)T_{lki}^{cba}R_j^d} \\ \nonumber + \underline{(kd\widetilde|lc)T_{lki}^{bca}R_j^d} + \underline{(kc\widetilde|ld)T_{kij}^{acd}R_l^b} + \underline{(kc\widetilde|ld)T_{ikl}^{cab}R_j^d}\Big].
\end{align}
All terms in the first matrix element, as well as the terms involving the Fock matrix in the second matrix element, scale as $O^3V^3$. First four terms in the third matrix element also possess $O^3V^3$ scaling, in the leading order, when we first contract the two-electron integrals with the single-excitation excited-state amplitudes. The remaining, underlined terms have either $O^4V^3$ (straight line) or $O^3V^4$ (wavy line) scaling. Now, we turn our attention to the triple amplitudes equation:
   \begin{align} \label{fourthmatrixel}
 \langle_{ijk}^{abc}| [ \widetilde{F}_N , R_{3} ]| 0\rangle = \frac{1}{2}P_3 \Big[ \uwave{\tilde{F}_{ad} R_{ijk}^{dbc}} - 
 \underline{\tilde{F}_{li} R_{ljk}^{abc}}
 \Big],
\end{align} 
\begin{align} \label{fifthmatrixel}
    \langle_{ijk}^{abc}| [ \tilde{H}_N , R_{2} ] |0\rangle &= P_3 \Big[\uwave{(bd\widetilde|ai)R_{jk}^{dc}} - \underline{(lj\widetilde|ai)R_{lk}^{bc}}\Big], 
\end{align}
\begin{align} \label{sixthmatrixel}
  \langle_{ijk}^{abc}| [ [ \tilde{H}_N , T_{2} ] , R_{1}] |0\rangle &= P_3 \Big[\underline{(mj\widetilde|li) T_{lk}^{ac} R_{m}^{b}} - \underline{(ai\widetilde|ld) T_{jk}^{bd} R_{l}^{c}} - \underline{(bd\widetilde|li) T_{jk}^{dc} R_{l}^{a}} \\ \nonumber
&  - \underline{(ai\widetilde|ld) T_{jl}^{bc}R_{k}^{d}} - \underline{(li\widetilde|cd) T_{lj}^{ab} R_{k}^{d}} + \uwave{(ad\widetilde|ce) T_{ji}^{bd} R_{k}^{e}} \Big]. \nonumber  
\end{align}
Each term in the above matrix elements has either $O^4V^3$ or $O^3 V^4$ scaling in the rate-limiting step, underlined according to the same convention as above. Thus, the overall scaling of the computational costs of the EOM-CC3 method can be characterized more precisely as $O^3 V^4$ in the leading order.

\section{Theory}

\subsection{Overview}
\label{sec:overview}

In this section we provide working equations of the rank-reduced equation-of-motion coupled cluster triples method (RR-EOM-CC3). In this method, the ground-state and excited-state triple-excitation amplitudes are expressed in the Tucker format (see the next section), while single and double excitations are treated in the conventional way (both in the ground-state and excited-state calculations). In our presentation of the RR-EOM-CC3 method, we omit terms appearing in the EOM-CCSD theory, denoted $O_1^{\mathrm{CCSD}}$ and $O_2^{\mathrm{CCSD}}$ above. They are known to have at most $N^6$ scaling and are evaluated using the conventional algorithm. For the remaining terms, we write out the factorized equations, so that the $N^6$ formal scaling of the developed method is evident. 

\subsection{Treatment of the $T_{ijk}^{abc}$ and $R_{ijk}^{abc}$ amplitudes}
\label{sec:tucker}

In this section we discuss general principles and workflow of the RR-EOM-CC3 theory.
First, both the ground-state and excited-state triple-excitation amplitudes are expressed in the Tucker-3 format:
\begin{align} 
    T_{ijk}^{abc} = t_{xyz}\,U_{ia}^x\,U_{jb}^y\,U_{kc}^z,  \label{Tucker1} \\
    R_{ijk}^{abc} = r_{XYZ}\,V_{ia}^X\,V_{jb}^Y\,V_{kc}^Z. \label{Tucker2}
\end{align}
Further in the text, the length of summations over $x,y,z$, i.e. dimension of the compressed subspace of the triply-excited ground-state amplitudes, is denoted by the symbol $N_{\mathrm{svd}}$. Similarly, the length of summations over $X,Y,Z$, i.e. dimension of the compressed subspace of the triply-excited excited-state amplitudes, is denoted by the symbol $N_{\mathrm{SVD}}$. Note that, when discussing scaling of particular terms later in the paper, we assume the following inequalities: $O \ll V < N_{\mathrm{SVD}} \approx N_{\mathrm{svd}} \ll N_{\mathrm{aux}}$.

We stress that in all calculations performed as part of the RR-EOM-CC3 routine, only the compressed amplitudes $t_{xyz}$ and $r_{XYZ}$, along with the tensors $U_{ia}^x$ and $V_{ia}^X$ that span the respective triple-excitation subspaces, are needed. For brevity, further in the text we refer to the quantities $U_{ia}^x$ and $V_{ia}^X$ as projectors. The full-rank amplitudes $T_{ijk}^{abc}$ and $R_{ijk}^{abc}$ are never explicitly calculated (even in batches) and stored in the rank-reduced approach. The projectors are found before the CC3/EOM-CC3 iterations by decomposition of some approximate triply-excited amplitudes and are fixed thereafter. The choices of approximate formulas for the amplitudes $T_{ijk}^{abc}$ and $R_{ijk}^{abc}$ are discussed below. Once the projectors are determined, the compressed amplitudes $t_{xyz}$ and $r_{XYZ}$ are found as solutions of the CC3 and EOM-CC3 equations within the subspace spanned by $U_{ia}^x$ and $V_{ia}^X$, respectively.

This leaves us with the choice of approximate amplitudes necessary to find the projectors and a method capable of efficient calculation of the decomposition given in Eqs.~\ref{Tucker1}~and~\ref{Tucker2}. Regarding the first point, for the ground-state CC3 calculation we follow the approach introduced in Refs.~\onlinecite{lesiuk2019,Lesiuk2020, Lesiuk2022} and use approximate $T_{ijk}^{abc}$ amplitudes that appear in (T) or [T] calculations. It was shown that this choice yields satisfactory accuracy both in the rank-reduced CC3~\cite{lesiuk2019} and rank-reduced CCSDT~\cite{Lesiuk2020} calculations for the ground state. For excited states, the choice of approximate $R_{ijk}^{abc}$ amplitudes has not been discussed thus far in the literature. We propose to employ the following formula:
\begin{equation}
\label{r3}
R_{ijk}^{abc} = (\epsilon_{ijk}^{abc} - \omega^{\mathrm{CCSD}} )^{-1} \langle _{ijk}^{abc}| [\tilde{V}_N , R_2^{\mathrm{CCSD}}] |0\rangle =   \frac{P_3\left[(ai\widetilde|lj)R_{lk}^{bc}-(ai\widetilde|bd)R_{jk}^{dc}\right]}{\epsilon_a + \epsilon_b + \epsilon_c - \epsilon_i - \epsilon_j - \epsilon_k - \omega^{\mathrm{CCSD}}},
\end{equation}
where the excitation energy $\omega^{\mathrm{CCSD}}$ and doubly-excited amplitudes $R_{ij}^{ab}$ come from the EOM-CCSD calculations which has been indicated by the superscripts. There are several ways to justify the adoption of this expression, but we focus on a purely theoretical argument based on L\"owdin partitioning~\cite{Löwdin1962, Löwdin1963} of the effective EOM Hamiltonian. This approach enables to calculate perturbative corrections to the EOM-CCSD energy taking EOM-CCSD as the reference (zeroth-order) wavefunction and is more consistent than M\"oller-Plesset partitioning~\cite{moller34} where the Hartree-Fock determinant is used as a reference. One can show that within the L\"owdin partitioning framework, the amplitudes from Eq.~\ref{r3} appear in the leading-order correction to the EOM-CCSD wavefunction due to the missing triple excitations~\cite{matthews16}, justifying their choice in the present context.

Of course, one can also criticize the choice given in Eq.~\ref{r3} using the argument that it is based solely on the information contained in the EOM-CCSD wavefunction. Therefore, one may expect that when the EOM-CCSD theory fails completely or yields unacceptably large errors, the guess based on Eq.~\ref{r3} becomes unsuitable. Such situation is encountered, e.g. for excited states with strong multireference character or for excited states dominated by double excitations -- EOM-CCSD errors of the order of several eV are not uncommon in these problematic cases. Fortunately, the numerical results given in the next section show that these objections are not warranted and the RR-EOM-CC3 method closely reproduces the canonical EOM-CC3 results also in situations where the EOM-CCSD excitation energies are not even qualitatively correct.

Next, we discuss how the Tucker decomposition of the amplitudes from Eq.~\ref{r3} is calculated without explicit construction of the full-rank amplitudes (which would incur $N^7$ cost). Similarly as for the ground-state amplitudes, we employ the higher-order orthogonal iteration (HOOI) procedure~\cite{kolda09,lathauwer2000}. A detailed description of this method in the context of the coupled cluster theory is given in Ref.~\onlinecite{Lesiuk2022} -- here we recall only its most salient features and describe changes necessary to apply it to excited-state amplitudes given by Eq.~\ref{r3}. The HOOI method can be viewed as a least-square minimization of the error of the decomposition in Eqs.~\ref{Tucker1}~and~\ref{Tucker2} for a fixed value of $N_{\mathrm{svd}}$/$N_{\mathrm{SVD}}$. The procedure is iterative in nature -- let us denote the projectors obtained in $n$-th iteration by $^{[n]}V_{ia}^X$. The key step of each HOOI iteration is calculation of the partially-projected quantity:
\begin{align}
    R_{ia,YZ}^{[n]} = R_{ijk}^{abc} \;^{[n]}V_{jb}^Y\,^{[n]}V_{kc}^Z.
\end{align}
The updated projectors, $^{[n+1]}V_{ia}^X$, are found by computing left singular vectors (using the singular-value decomposition) of $R_{ia,YZ}^{[n]}$ that correspond to the largest singular values. This procedure is repeated until the decomposition error no longer decreases. It is important to point out that as a side-effect of this algorithm, the converged projectors are orthonormal in the sense that $V_{ia}^X\,V_{ia}^Y = \delta_{XY}$ (summation over the indices $ia$ is implied). This virtue of the projectors is useful in simplifying the working equations and shall be used further in the text.

The robustness of the HOOI procedure stems from the fact that only the quantity $R_{ia,YZ}^{[n]}$ is referenced and the full-rank amplitudes are never required. In Ref.~\onlinecite{Lesiuk2022} it has been shown that by using the numerical Laplace transform of the energy denominator (which introduces negligible errors), the costs of calculating $R_{ia,YZ}^{[n]}$ are proportional to $N^5$ in the case of ground-state amplitudes. This procedure is straightforward to adapt to the excited-state amplitudes -- the only changes are the replacement of the ground-state amplitudes, $T_2$, by their excited-state counterparts, $R_2$, and the inclusion of the shift by $\omega^{\mathrm{CCSD}}$ in the denominator. Neither change affects the cost of the calculations in a meaningful way.

Despite no fundamental difficulties in application of the HOOI method to $R_{ijk}^{abc}$, we have encountered a minor technical problem which was absent (or at least extremely rare) in ground-state calculations. Namely, in some systems the HOOI iterations oscillate between two states, $^{[n]}V_{ia}^X$ and $^{[n+1]}V_{ia}^X$, which do not constitute a satisfactory solution. In the context of, e.g. self-consistent field iterations, the standard approach to quench such oscillations (referred to as "damping") is to take a linear combination of the two solutions in the next iteration. Unfortunately, this straightforward damping amounting to, e.g. averaging $^{[n]}V_{ia}^X$ and $^{[n+1]}V_{ia}^X$, is not permissible in HOOI for several reasons -- the most important of which is that the aforementioned orthonormality of the projectors would be violated. We propose a slightly more advanced procedure to perform the damping in the HOOI iterations. Having the projectors from two consecutive iterations, $^{[n]}V_{ia}^X$ and $^{[n+1]}V_{ia}^X$, both with dimensions $OV\times N_{\mathrm{SVD}}$, we stack them together and temporarily form an object $W_{ia}^X$ of dimension $OV\times 2N_{\mathrm{SVD}}$. Next, we perform the singular-value decomposition of $W_{ia}^X$ and as the next projectors we take the left singular vectors corresponding to the largest singular values. In our experience, this simple and inexpensive modification of the HOOI procedure alleviates the oscillation issue encountered in applications to the excited-state amplitudes. Note that this procedure can be extended further by including additional projectors from previous iterations, $^{[n-1]}V_{ia}^X$, $^{[n]}V_{ia}^X$ and $^{[n+1]}V_{ia}^X$, etc., akin to the well-known DIIS acceleration. However, this is not necessary from the point of view of the present work.

\subsection{Triple amplitudes equation}
\label{sec:tripeq}

An advantage of the EOM-CC3 method is that the triple amplitudes equation is considerably simplified in comparison with EOM-CCSDT and the expression for the full-rank triply-excited amplitudes can be written explicitly. In fact, starting with equation \ref{tripleeq} and adopting the usual approximation used in the CC3/EOM-CC3 theory
\begin{align} \label{Fock}
 \langle_{ijk}^{abc}| [ \tilde{F}_N , R_{3} ]| 0\rangle \approx ( \epsilon_{a} + \epsilon_{b} + \epsilon_{c} - \epsilon_{i} - \epsilon_{j} - \epsilon_{k})  R_{ijk}^{abc},
\end{align}
we obtain
\begin{align} 
\label{defr3}
 ( \epsilon_{a} + \epsilon_{b} + \epsilon_{c} - \epsilon_{i} - \epsilon_{j} - \epsilon_{k} - \omega)  R_{ijk}^{abc} = - \langle_{ijk}^{abc}| \big[ [ \tilde{H}_N , T_{2} ] , R_{1}\big] + [ \tilde{H}_N , R_{2} ] | 0\rangle,
\end{align}
where the right-hand side does not depend on $R_{ijk}^{abc}$. Our first goal is to find an analogous expression for the compressed amplitudes $r_{XYZ}$.

For brevity, we denote the right-hand side of the above equation as $P_3\gamma_{ijk}^{abc}$, where $P_3$ is the permutation operator defined in the previous section. The explicit form of $\gamma_{ijk}^{abc}$ can be found by summing the right-hand sides of Eqs. \ref{fifthmatrixel} and \ref{sixthmatrixel}, reversing signs and removing the permutation operator $P_3$, leading to
\begin{align} 
\label{explicitgammawhynot}
\begin{split}
    &\gamma_{ijk}^{abc} = 
    (lj\widetilde|ai)R_{lk}^{bc} - 
    (bd\widetilde|ai)R_{jk}^{dc} - 
    (mj\widetilde|li) T_{lk}^{ac} R_{m}^{b} + 
    (ai\widetilde|ld) T_{jk}^{bd} R_{l}^{c} \\
    &+
    (bd\widetilde|li) T_{jk}^{dc} R_{l}^{a} + 
    (ai\widetilde|ld) T_{jl}^{bc} R_{k}^{d} + 
    (li\widetilde|cd) T_{lj}^{ab} R_{k}^{d} - 
    (ad\widetilde|ce) T_{ji}^{bd} R_{k}^{e}.
\end{split}
\end{align}
Inserting the Tucker decomposition of the triple-excitation amplitudes $R_{ijk}^{abc}$ into Eq.~\ref{defr3} and using the definition of $\gamma_{ijk}^{abc}$ one finds:
\begin{align} \label{}
 ( \epsilon_{a} + \epsilon_{b} + \epsilon_{c} - \epsilon_{i} - \epsilon_{j} - \epsilon_{k} - \omega)\, r_{XYZ}\,V_{ia}^X\,V_{jb}^Y\,V_{kc}^Z = P_3\,\gamma_{ijk}^{abc}.
\end{align}
 Note that in the above equation we  sum only over $X$, $Y$ and $Z$ indices. Next, we project this equation onto the triple-excitation subspace spanned by the expansion tensors $V_{ia}^X$. This is done by multiplying both sides by $V_{ia}^{X'} V_{jb}^{Y'} V_{kc}^{Z'}$ and performing summations over indices $i,j,k,a,b,c$. As a result we get:
\begin{align} \label{triplecore}
 r_{XYZ}\,V_{ia}^X\,V_{jb}^Y\,V_{kc}^Z \big[ (\epsilon_{a} - \epsilon_{i}) + (\epsilon_{b} - \epsilon_{j}) + (\epsilon_{c} - \epsilon_{k}) - \omega \big]   V_{ia}^{X'} V_{jb}^{Y'} V_{kc}^{Z'} =  V_{ia}^{X'}V_{jb}^{Y'} V_{kc}^{Z'}P_3\,\gamma_{ijk}^{abc}.
\end{align}
This equation can be considerably simplified if the following two conditions are enforced
\begin{align}
 V_{ia}^{X}(\epsilon_{a} - \epsilon_{i})V_{ia}^{X'} &=
 \epsilon_{X}\,\delta_{XX'},\\
 V_{ia}^{X}V_{ia}^{X'} &= \delta_{XX'}. 
\end{align}
As mentioned in Sec.~\ref{sec:tucker}, the second condition (orthonormality) is a natural consequence of the HOOI procedure used to find $V_{ia}^X$. The first condition can also be imposed without loss of generality by performing an orthogonal transformation $O_{XY}$ of the original basis vectors, i.e.~$V_{ia}^{Y} \leftarrow V_{ia}^{X} O_{XY}$, thereby eliminating the "rotational" ambiguity in the choice of the basis (at least in cases where all $\epsilon_{X}$ are distinct). To sum up, both of the above conditions can be enforced upon the projectors without changing the span of the triple-excitation subspace itself. This leads to the following simplified expression for the compressed amplitudes
\begin{align}
 ({\epsilon_{X} + \epsilon_{Y} + \epsilon_{Z} - \omega})\,r_{XYZ} = {V_{ia}^{X}\,V_{jb}^{Y}\,V_{kc}^{Z}\big(P_3 \gamma_{ijk}^{abc}\big)}.
\end{align}
In the last step, we replace the permutation over indices $i,j,k,a,b,c$ with the permutation over $X,Y,Z$ which is possible due to general relation:
\begin{align}
\begin{split}
    &V_{ia}^{X}\,V_{jb}^{Y}\,V_{kc}^{Z}\big(P_3 A_{ijk}^{abc}\big) = P_{XYZ}\big(V_{ia}^{X}\,V_{jb}^{Y}\,V_{kc}^{Z}\,A_{ijk}^{abc}\big), \\
    &P_{XYZ} = (1 + P_{XY})(1 + P_{XZ} + P_{YZ}),
\end{split}
\end{align}
which holds for an arbitrary tensor $A_{ijk}^{abc}$. Finally, the expression for the core tensor $r_{XYZ}$ takes the form: 
\begin{align} \label{rXYZ}
r_{XYZ} = \frac{P_{XYZ}\big(V_{ia}^{X}\,V_{jb}^{Y}\,V_{kc}^{Z}\,\gamma_{ijk}^{abc}\big)}{\epsilon_{X} + \epsilon_{Y} + \epsilon_{Z} - \omega}.
\end{align}
This equation can be factorized, so that the evaluation of $r_{XYZ}$ requires no steps scaling higher than $N^5$ with the system size. In order to show this fact we need to introduce a couple of intermediates:
\begin{align} 
 &\tau_{la}^{Z} = T_{lk}^{ac} V_{kc}^{Z}, \;\;\;
 \chi^{QX} = B_{ai}^{Q} V_{ia}^{X}, \;\;\;
 \chi_{ld}^{X} = \chi^{QX} B_{ld}^{Q},
\end{align}
\begin{align} 
 &\xi_{li}^{Y} = B_{li}^{Q} \left(B_{mj}^{Q} (V_{jb}^{Y} R_{m}^{b})\right),
 &&\tau_{li}^{Y} = B_{li}^{Q} \left(B_{cd}^{Q} (V_{kc}^{Y} R_{k}^{d})\right), \\
 &\xi_{ad}^{Z} = B_{ad}^{Q} \left(V_{kc}^{Z} ( B_{ce}^{Q} R_{k}^{e})\right),
 &&\tau_{bd}^{X} = B_{bd}^{Q} \left(V_{ia}^{X} ( B_{li}^{Q} R_{l}^{a})\right), \\
 &\xi^{QYZ} = \left(B_{lj}^{Q} (V_{kc}^{Z} R_{lk}^{bc})\right)  V_{jb}^{Y},
 &&\tau^{QYZ} = \left(B_{bd}^{Q} (V_{kc}^{Z} R_{jk}^{dc})\right)  V_{jb}^{Y}.
\end{align}
The parentheses in the above equations indicate the optimal order of operations. As can be seen, there are no intermediates with higher than $N^5$ formal scaling. The most expensive are $\xi^{QYZ}$ and $\tau^{QYZ}$ which both scale as $O V N_{\mathrm{SVD}}^2 N_{\mathrm{aux}}$ in the leading order. The $\tau_{la}^{Z}$ intermediate is less costly, scaling as $O^2 V^2 N_{\mathrm{SVD}}$, while the remaining intermediates scale only with the fourth power of the system size. With help of the above intermediates and the explicit form of $\gamma_{ijk}^{abc}$ given in Eq.~\ref{explicitgammawhynot}, we rewrite Eq.~\ref{rXYZ} in fully explicit form

\begin{align} \label{r}
 r_{XYZ} &= \frac{P_{XYZ}  }{\epsilon_{X} + \epsilon_{Y} + \epsilon_{Z} - \omega} \left[(\tau_{li}^{Y} - \xi_{li}^{Y}) \left( \tau_{la}^{Z}\,V_{ia}^{X} \right) + ( \chi_{ld}^{X}\,\tau_{kd}^{Y} ) ( R_{l}^{c}\,V_{kc}^{Z} ) \right. \nonumber \\ \nonumber \\
 & \left. + ( \chi_{ld}^{X}\,R_{k}^{d} ) ( \tau_{lc}^{Y}\,V_{kc}^{Z}) + \left( \tau_{bd}^{X}\,\tau_{jd}^{Z} \right) V_{jb}^{Y} - \left( \xi_{ad}^{Z}\,\tau_{id}^{Y}\, \right) V_{ia}^{X} + \chi^{QX} (\xi^{QYZ} - \tau^{QYZ}) \right].
 \end{align}
 Again, no single step in the above equation involves more than five unique indices. The most expensive are the fourth and the fifth terms with scaling of $O V N_{\mathrm{SVD}}^3$ in the leading order. Clearly, in the rank-reduced formalism, the determination of the compressed amplitudes tensor $r_{XYZ}$ becomes a relatively inexpensive task with $N^5$ scaling.
 
 \subsection{Triples contributions to single and double amplitudes equations}
\label{sec:eomsd}
 
 We now proceed to the factorization of equations resulting from projections onto singly- and doubly-excited determinants. The former projection involves only one term, $\langle_{i}^{a}| [ \tilde{H}_N , R_{3} ] |0\rangle$, given by Eq.~\ref{firstmatrixel} with full-rank triply-excited amplitudes. If the Tucker format is used for $R_{ijk}^{abc}$, together with the following intermediates
 \begin{align} \label{intA}
 A_{X}^{Q} = B_{ia}^{Q}\,V_{ia}^{X},
 \\ \label{intB}
 B_{ij}^{QX} = B_{ia}^{Q}\,V_{ja}^{X},
\end{align}
this term can be written in the following form:
\begin{align} 
\label{firstel}
\langle_{i}^{a}| [ \tilde{H}_N , R_{3} ] |0\rangle &= \left[ \big(2A_{Y}^{Q}\,A_{Z}^{Q} - B_{jk}^{QZ}B_{kj}^{QY}
\big)\,r_{XYZ} \right]V_{ia}^{X} 
\nonumber \\ 
&- \left[ \big(2 A_{Z}^{Q}\,r_{XYZ}\big) B_{ji}^{QX} - \big(B_{jk}^{QZ}\,B_{ki}^{QX}\big)\,r_{XYZ}\right]V_{ja}^{Y}.
\end{align}
The last term in the second square brackets in the above equation is the most computationally expensive. The first step of its evaluation involves six unique indices, so the formal scaling is $N^6$ or, more precisely, $O^3 N_{\mathrm{SVD}}^2 N_{\mathrm{aux}}$.

Next, we consider the projection onto doubly-excited determinants. We have to deal with two unique terms, namely $\langle_{ij}^{ab}| [ \tilde{H}_N , R_{3} ] |0\rangle$ and $\langle_{ij}^{ab}| \big[[ \tilde{H}_N , T_{3} ], R_1 \big]|0\rangle$, see Eqs.~\ref{secondmatrixel}~and~\ref{thirdmatrixel}. We divide the first term into two parts as follows:
\begin{equation}
\langle_{ij}^{ab}| [ \tilde{H}_N , R_{3} ] |0\rangle = \langle_{ij}^{ab}| [ \widetilde{F}_N , R_{3} ] |0\rangle + \langle_{ij}^{ab}| [ \widetilde{V}_N , R_{3} ] |0\rangle.
\end{equation}
The factorization of the part involving the Fock operator is straightforward:
\begin{align} \label{secondelF}
\begin{split}
\langle_{ij}^{ab}| [ \widetilde{F}_N , R_{3} ] |0\rangle &= 
P_2\Big[\big[\big((\widetilde{F}_{kc}\,V_{kc}^{Z})\,r_{XYZ}\big)V_{jb}^{Y} - \left((\widetilde{F}_{kc}\,V_{jc}^{Y})\,r_{XYZ}\right)V_{kb}^{Z}\big]V_{ia}^{X}\Big].
\end{split}
\end{align}
This part has the overall $N^5$ scaling. The second part is more complicated and several intermediates have to be introduced to achieve an optimal factorization, namely
\begin{align}
\Pi^{QYZ} = A_{X}^{Q}\,r_{XYZ}, \;\;\;
\Pi_{ik}^{X} = B_{li}^{Q}\,B_{kl}^{QX}, \;\;\;
\Pi_{ia}^{QZ} = B_{ki}^{Q}\,V_{ka}^{Z},
\end{align}
\begin{align}
\Gamma_{ia}^{YZ} = V_{ia}^{X}\,r_{XYZ}, \;\;\;
\Gamma_{ia}^{QZ} = B_{ad}^{Q}\,V_{id}^{Z}.
\end{align}
In the above, the intermediates $A_{X}^{Q}$ and $B_{kl}^{QX}$ are the same as in Eqs.~\ref{intA}~and~\ref{intB}. Factorization yields the following expression for the second part:
\begin{align}
\label{projVR3}
    \langle_{ij}^{ab}| [ \widetilde{V}_N , R_{3} ] |0\rangle = P_2 \big[ \big(V_{ka}^{Y} \Pi_{ik}^{Z} - \Gamma_{ka}^{QZ}B_{ki}^{QY}
    \big)\Gamma_{jb}^{YZ} + \big(\Gamma_{ia}^{QZ} - \Pi_{ia}^{QZ}\big)\big(2\Pi^{QYZ}V_{jb}^{Y} - \Gamma_{kb}^{YZ}B_{kj}^{QY}\big)   \big].
\end{align}
Neither the construction of intermediates nor the evaluation of the above formula requires any $N^7$ steps. The most expensive contractions here are $\Gamma_{ka}^{QZ}B_{ki}^{QY}$ and $\Gamma_{kb}^{YZ}B_{kj}^{QY}$, both having $O^2V N_{\mathrm{SVD}}^2 N_{\mathrm{aux}}$ scaling. Additionally, contraction between the last two parentheses scales as $O^2V^2 N_{\mathrm{SVD}} N_{\mathrm{aux}}$.

Lastly, we turn our attention to the remaining matrix element, $\langle_{ij}^{ab}| \big[[ \tilde{H}_N , T_{3} ], R_1 \big]|0\rangle$. We introduce a handful of new intermediates:
\begin{align}
&(kc\widetilde{\widetilde|}ld) = 2(kc\widetilde|ld) - (kd\widetilde|lc), \\
&\alpha_{ld}^x = (kc\widetilde{\widetilde|}ld)\,U_{kc}^x, \;\;\;
\beta_{kc} = (kc\widetilde{\widetilde|}ld)R_l^d,\\
&\gamma_{iald}^z = U_{ic}^y\,t_{xyz}\,U_{ka}^x\,B_{kc}^Q\,B_{ld}^Q.
\end{align}
Then, the formula for the required matrix element can be rewritten as:
\begin{align}
\label{proj22}
\langle_{ij}^{ab}| \big[[ \tilde{H}_N , T_{3} ], R_1 \big]|0\rangle =
P_2 \big[ \big(\alpha_{ld}^x \, R_l^d \,
t_{xyz} \, U_{jb}^y - \beta_{kc} \, U_{jc}^y \,
t_{xyz} \, U_{kb}^x - \alpha_{ld}^x \, U_{jd}^y \, t_{xyz} \, R_l^b
 \nonumber \\  - \alpha_{ld}^x \, R_j^d \, t_{xyz} \, U_{lb}^y
\big)\,U_{ia}^z + \gamma_{iald}^z \, U_{jd}^z \, R_l^b + \gamma_{iald}^z \, R_j^d \, U_{lb}^z\big].
\end{align}
We evaluate each term in the above matrix element and the $\gamma_{iald}^z$ intermediate via step-by-step contraction from left to right. The most expensive step in the calculation of $\langle_{ij}^{ab}| \big[[ \tilde{H}_N , T_{3} ], R_1 \big]|0\rangle$ is the formation of the $\gamma_{iald}^z$ intermediate. This procedure costs $2O^2V^2 N_{\mathrm{svd}} N_{\mathrm{aux}}$ in the leading order. As a technical note, in our implementation the intermediate $\gamma_{iald}^z$ is formed in batched loops over $z$ and consumed immediately in order to reduce memory requirements.

Equations presented in this section, along with the factorized equation for the compressed triple-excitation amplitudes $r_{XYZ}$ -- Eq.~\ref{r}, are a proof of the overall $N^6$  scaling of the RR-EOM-CC3 method. Moreover, we expect the evaluation of Eqs.~\ref{projVR3}~and~\ref{proj22} to be rate-limiting for large systems due to the cost $2O^2 V N_{\mathrm{SVD}}^2 N_{\mathrm{aux}}+O^2 V^2 N_{\mathrm{SVD}} N_{\mathrm{aux}}$ and $2O^2V^2 N_{\mathrm{svd}} N_{\mathrm{aux}}$, respectively. It is also useful to compare this scaling with the canonical EOM-CC3, where the overall computational cost scales as $8O^3V^4$, see Ref.~\onlinecite{Paul2021} for a detailed analysis. To enable a meaningful comparison, we estimate that in most practical situations $N_{\mathrm{svd}}\approx N_{\mathrm{SVD}}\approx V$. This rough estimate is justified by numerical results given in the subsequent sections. Next, the size of the auxiliary basis set is usually larger by a factor $3-4$ than the orbital basis set or, to simplify the analysis, the number of virtual orbitals, $V$. By taking the ratio $2O^2 V N_{\mathrm{SVD}}^2 N_{\mathrm{aux}}+O^2 V^2 N_{\mathrm{SVD}} N_{\mathrm{aux}}+2O^2V^2 N_{\mathrm{svd}} N_{\mathrm{aux}}$ to $8O^3V^4$ and using these estimates for $N_{\mathrm{svd}}$, $N_{\mathrm{SVD}}$ and $N_{\mathrm{aux}}$, we find that the rank-reduced formalism becomes beneficial for $O>2-3$. Of course, this analysis does not take into account that in the rank-reduced formalism EOM-CCSD ($O^2V^4$ scaling) and HOOI ($N^5$ scaling) calculations have to be performed before the EOM-CC3 procedure. While the scaling of HOOI algorithm is low it has a large prefactor and still contributes to the overall computational costs for systems that are currently within reach. Nonetheless, this estimate suggests that the rank-reduced formalism should possess a relatively sudden cross-over point with the canonical EOM-CC3 in terms of computational costs.

\section{Results and discussion}

\subsection{Computational Details}
\label{sec:comput}

We tested the accuracy of the RR-EOM-CC3 approach by comparing it to the exact EOM-CC3 results in a series of calculations of vertical excitation energies. The calculations can be divided into three distinct parts: benchmarks for small molecules, applications to moderate/large molecules and tests for states with pronounced doubly-excited character. All calculations have been performed in augmented Dunning-type basis sets aug-cc-pVDZ and aug-cc-pVTZ \cite{dunning1989a, kendall1992a}. In the calculations for larger molecules, density fitting approximation of two-electron integrals was utilized. The standard aug-cc-pVXZ-RI auxiliary basis sets were used for this purpose~\cite{weigend2002}. Moreover, frozen-core approximation was used throughout -- $1s$ orbitals of first-row atoms were not correlated.

The geometries of small molecules (listed below) have been optimized at the B3LYP-D3/cc-pVTZ~\cite{dunning1989a, Becke1993, Stephens1994, Grimme} level of theory with PSI4~\cite{PSI4} quantum chemistry program. The optimized structures are given in the Supporting Information. In calculations for doubly-excited states we used high-quality geometries reported by Loos et. al. \cite{Loos2022}. The geometries of two large molecules investigated in this work, L-proline and heptazine, have also been taken from the literature.
For L-proline we use the same geometry as that utilized in the article by Paul et. al., who adopted it from the Pubchem database \cite{Paul2021, L-proline}. The heptazine geometry was in turn optimized by Loos et. al. and reported in the recent article \cite{Loos2023}.

The RR-EOM-CC3 routine relies on the Davidson's algorithm for calculation of excitation energies~\cite{davidson1975, Liu, butscher1976}. Block version of this algorithm is utilized in the EOM-CCSD calculation, providing a reasonable guess for EOM-CC3 which itself uses single-root version of the algorithm. In the present work, thresholds for convergence of the Davidson's algorithm were set to $10^{-5}$ hartree for excitation energy and $10^{-4}$ for the value of the maximum coefficient of a residual vector, both for block and single-root steps. However, there were three exceptions. The thresholds used for single-root (EOM-CC3) calculations of glyoxal and benzene in the aug-cc-pVTZ basis set were by an order of magnitude smaller than mentioned above (for both energy and residual vector). Another exception was heptazine molecule in aug-cc-pVTZ basis set where the threshold for a residual vector was set to $10^{-3.5}$ in both block and single-root steps of the routine. The loosening of thresholds in the case of heptazine was done primarily for reasons of computational efficiency and has no influence on the results within accuracy levels reported in this work.

\subsection{Small Molecules}
\label{sec:small}

In this section we investigate the accuracy of the RR-EOM-CC3 method when used in calculations for singly-excited states of small molecules. However, it is first desirable to discuss the significance of the two parameters of the developed method, namely, the sizes of the triple-excitation subspaces in the ground state ($N_{\mathrm{svd}}$) and in the excited state ($N_{\mathrm{SVD}}$). We note that the selection of the triple-excitation subspace sizes is an important choice in RR-EOM-CC3. It has a direct influence on the accuracy and computational costs of calculations. One way of investigating this influence is simply to perform calculations for a wide range of  $N_{\mathrm{SVD}}$ and $N_{\mathrm{svd}}$ values (see Fig.~S1 in Supporting Information). While this approach is informative, it might be preferable to relate the two subspace sizes to each other, so that we are effectively left with only one parameter. The benefits of this approach are twofold, namely, it simplifies the analysis of the results and gives a more straightforward method to the end user. Such relation must, however, fulfill two basic requirements. Firstly, it must result in a linear scaling of both parameters with the system size. Secondly, a systematic expansion of the subspaces must lead to the exact EOM-CC3 energy when the excitation subspace includes all possible excitations within given basis. Both requirements are met if we assume  a linear relation, $N_{\mathrm{SVD}} = a N_{\mathrm{svd}} + b$, where $a>0$ and $b$ are constants. In the following we focus on the "diagonal", i.e. $N_{\mathrm{SVD}} = N_{\mathrm{svd}}$. We emphasize that it is an arbitrary, although somewhat justified, choice. From the theoretical viewpoint, having the same subspace sizes for the ground state and the investigated excited state means that there is no inherent bias towards any particular state. This is important in calculations of excitation energies, where the result is a difference between the energies of the aforementioned states (even though the excitation energy is calculated directly in the EOM procedure).

\begin{figure}[H]
    \centering
    \includegraphics{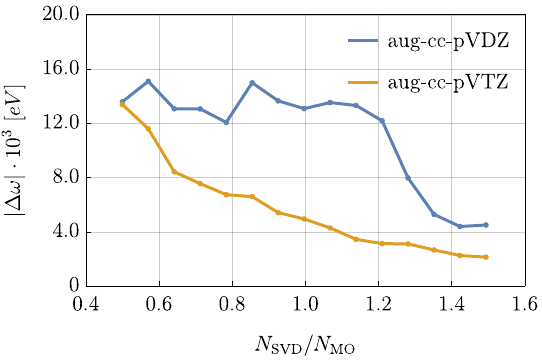}
    \caption{\footnotesize Absolute value of the difference in the excitation energy calculated with the rank-reduced and exact EOM-CC3 plotted against the ratio of the excitation subspace size ($N_{\mathrm{SVD}}$) and the number of molecular orbitals of a given molecule ($N_{\mathrm{MO}}$). The plot represents the results averaged over the chosen set of small molecules.}
    \label{average_error}
\end{figure}

With this in mind, we now turn our attention to the results obtained along the diagonal, with a single parameter  $N_{\mathrm{SVD}}$ denoting the triple-excitation subspace size for both the ground state and the excited state. We calculated the first excitation energies for a set of eight small molecules -- $\mathrm{BH}_3$, $\mathrm{C}_2\mathrm{H}_2$, $\mathrm{C}_2\mathrm{H}_4$, $\mathrm{CH}_3\mathrm{OH}$, $\mathrm{CO}$, $\mathrm{H}_2\mathrm{O}$, $\mathrm{CH}_2\mathrm{O}$, $\mathrm{NH}_3$ -- and compared them with the exact EOM-CC3 calculations carried out in the {\sc PSI4} program. The results are presented in Fig.~\ref{average_error}, which shows the absolute value of the error \big(|$\Delta \omega$|\big) between the RR-EOM-CC3 method and its exact counterpart in calculations of the first excitation energies of the investigated molecules as a function of the triple-excitation subspace size. The size of the subspace is given in the units of $N_{\mathrm{MO}}$, which is the number of molecular orbitals of a given system, excluding the frozen-core orbitals. The errors for each $N_{\mathrm{SVD}}$ are averaged over the chosen set of systems. 

From Fig.~\ref{average_error} it becomes clear that the RR-EOM-CC3 method performs particularly well in terms of accuracy in the triple-zeta basis. First, the errors decrease rather quickly and in a regular fashion as a function of the $N_{\mathrm{SVD}}$ parameter. Second, even for modest sizes of the triple excitation subspace ($N_{\mathrm{SVD}}\approx N_{\mathrm{MO}}$), the errors are of the order of $0.005\,$eV. Note that the inherent error of the EOM-CC3 method for singly-excited states (in comparison with FCI) equals 0.03 eV as reported by Loos et. al.~\cite{loos2018} This means that errors resulting from the applied decomposition of triply-excited amplitudes are several times smaller (for $N_{\mathrm{SVD}}\approx N_{\mathrm{MO}}$) than the inherent error of the exact EOM-CC3 excitation energy for this kind of states. In the case of the aug-cc-pVDZ basis, it is also possible to achieve this level of accuracy, but with somewhat larger sizes of the excitation subspace. Moreover, the convergence of the results as a function of $N_{\mathrm{SVD}}$ is less regular in the smaller basis and a mildly oscillatory behavior is observed for $N_{\mathrm{SVD}}< N_{\mathrm{MO}}$. The fact that the performance of the rank-reduced formalism improves in a larger basis is not entirely unexpected -- a similar behavior was found previously in ground-state calculations~\cite{Lesiuk2022}.

It is desirable to determine a recommended value of the $N_{\mathrm{SVD}}$ parameter for both basis sets. In the case of aug-cc-pVTZ the choice of $N_{\mathrm{SVD}} = N_{\mathrm{MO}}$ seems reasonable, as it gives an error which is six times smaller than the inherent error of EOM-CC3 for the singly-excited states. From the pragmatic point of view of an end user, the RR-EOM-CC3 method would be indistinguishable from the canonical EOM-CC3 in this regime. The recommended triple-excitation subspace size for aug-cc-pVDZ must be larger if one wants to maintain a similar level of accuracy. In this case a sensible choice is $N_{\mathrm{SVD}}$ between $1.35N_{\mathrm{MO}}$ and $1.4N_{\mathrm{MO}}$, although a smaller subspace size may be enough in many applications. One has to keep in mind that in the aug-cc-pVDZ basis, the basis set incompleteness error is likely to be substantial and comparable to the error of the method itself. In such situations, increasing the accuracy of the rank-reduced formalism to a level well below $0.01\,$eV would not improve the overall quality of the results and may not be necessary.

\begin{figure}[H]
    \centering
    \includegraphics{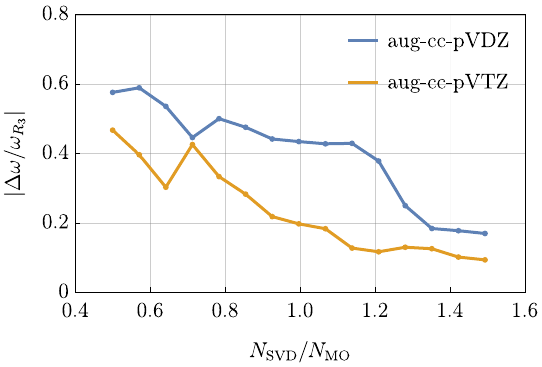}
    \caption{\footnotesize Absolute value of the error in RR-EOM-CC3 divided by the absolute value of the exact EOM-CC3 triple amplitudes contribution to the excitation energy plotted against the ratio of the decomposition basis size ($N_{\mathrm{SVD}}$) and number of molecular orbitals of a given molecule ($N_{\mathrm{MO}}$). The plot represents the results averaged over the chosen set of small molecules.}
    \label{average_tryp}
\end{figure}

In the previous paragraphs we discussed the average errors in the total RR-EOM-CC3 excitation energy. However, as the difference between the EOM-CCSD and EOM-CC3 results is not large for states dominated by single excitations, it is advisable to compare the error of the RR-EOM-CC3 method with the difference between the exact EOM-CC3 and EOM-CCSD excitation energies. The latter quantity is denoted by the symbol $\omega_{R_3}$ as it quantifies the contribution of triply-excited configurations to the excitation energy. Then the ratio |$\Delta \omega/\omega_{R_3}$| tells us whether the RR-EOM-CC3 method performs better in terms of accuracy than EOM-CCSD for a given molecule. Clearly, this is a much more demanding test than comparing just the total excitation energy. The values of the quantity |$\Delta \omega/\omega_{R_3}$| averaged over the chosen set of small molecules as a function of triple-excitation subspace size are presented in Fig.~\ref{average_tryp}.

From Fig.~\ref{average_tryp} we note that, again, the aug-cc-pVTZ basis set gives better results than aug-cc-pVDZ. For $N_\mathrm{{SVD}} = N_{\mathrm{MO}}$ the investigated parameter equals approximately $0.2$ in the former basis set and decreases further for larger triple-excitation subspace sizes. For the aug-cc-pVDZ basis set, the ratio $|\Delta \omega/\omega_{R_3}|$ is between 0.4 and 0.6 in small and moderate subspace sizes and decreases to lower values at around $N_\mathrm{{SVD}} = 1.2N_{\mathrm{MO}}$. Although the error $|\Delta \omega|$ reaches at some points over half of the value of $|\omega_{R_3}|$ we note that the latter quantity is rather small for the investigated molecules. When averaged over all systems, it equals 0.045 eV and 0.042 eV in aug-cc-pVDZ and aug-cc-pVTZ basis sets, respectively. Furthermore, we emphasize that increasing $N_{\mathrm{SVD}}$ beyond $1.5 N_{\mathrm{MO}}$ still should result in a reasonable compression of the triple-excitation amplitudes. This is because the maximal value of $N_{\mathrm{SVD}}$ in a given basis is equal to the product of the number of occupied and virtual orbitals, $OV$. Thus, the accuracy of the RR-EOM-CC3 method can be further increased if needed.  

\subsection{Doubly-excited states}
\label{sec:doubly}

In the second part of the calculations we investigated isolated states with doubly-excited character for each of the following molecules: acrolein, butadiene, benzene, nitrosomethane, nitroxyl and glyoxal. The obtained RR-EOM-CC3 excitation energies were compared to the exact EOM-CC3 data taken from the article by Loos et. al. \cite{Loos2022} In each case, we set the ground-state and the excited-state triple-excitation subspace sizes equal to the number of molecular orbitals of a given molecule ($N_{\mathrm{svd}}$ = $N_{\mathrm{SVD}}$ = $N_{\mathrm{MO}}$). We note that this choice in the case of the aug-cc-pVDZ basis set was deemed suboptimal in the previous section. However, it turns out that the results obtained with this value of $N_{\mathrm{SVD}}$ are satisfactory for the molecules investigated here and further extension is not necessary. 

In Table~\ref{doubly-excited} we report results of the calculations for the doubly-excited states. Additionally, we show values of the $\%R_2$ parameter which can be used as an indicator to which degree a given state has a doubly-excited character. It is defined as the square norm of the $R_{ij}^{ab}$ amplitudes (the eigenvector is normalized to the unity in our implementation). We emphasize that the $\%R_2$ parameter is calculated here without a constraint to only unique excitations. In the remaining columns of Table~\ref{doubly-excited} we report excitation energies calculated with the RR-EOM-CC3 method, the exact EOM-CCSD (from the block Davidson step of our routine) and the exact EOM-CC3 (from the literature). The last two columns show the same quantities that we investigated for singly-excited states, namely, the RR-EOM-CC3 error ($\Delta \omega$) and this error divided by the triple amplitudes contribution to the energy ($\% |\Delta \omega / \omega_{R_3}|$). We note that there is a varying degree of double-excitation character ($\%R_2$) among the considered states. The states of the nitrosomethane, nitroxyl and glyoxal molecules are of pure double-excitation character, while the characters of the remaining states lie somewhere in between this extreme and the singly-excited states studied in the previous section.  

\begin{table}[H] 
\centering
\small
    \begin{tabular}{M{3cm}|M{1cm}|M{2cm}|S[table-format=1.3]|M{2cm}|S[table-format=-1.4]|S[table-format=2.2]}
    \hline
    \multicolumn{7}{M{14cm}}{aug-cc-pVDZ} \\
    \hline
        molecule   &$\%R_2$ & {RR-CC3} & {ex. CCSD} & {ex. CC3$^a$} & $\Delta \omega$ & $\%|\Delta \omega / \omega_{R_3}|$ \\
         \hline
         acrolein  &32.2 & 6.785 &7.200 &6.754 &0.031 & 7.1 \\
         butadiene  &42.9 & 6.687 &7.086 &6.678 &0.009 & 2.3 \\
         benzene  &23.9 & 5.107 &5.213 &5.114 & -0.007 & 7.2 \\
         nitrosomethane & 99.3 & 5.767 &9.081 &5.749 &0.018 & 0.53  \\
         nitroxyl   &99.9 & 5.264 &8.245 &5.247 & 0.017 & 0.56  \\
         glyoxal  &99.9 & 6.738 &11.575 &6.706 &0.032 & 0.67  \\
          \hline
    \multicolumn{7}{M{14cm}}{aug-cc-pVTZ} \\
    \hline
     molecule &$\%R_2$ & {RR-CC3} & {ex. CCSD} & {ex. CC3$^a$}& $\Delta \omega$ & $\%|\Delta \omega / \omega_{R_3}|$ \\
         \hline
         acrolein  &32.3 & 6.775 &7.269 &6.752 &0.023 & 4.5 \\
         butadiene  &42.7 & 6.682 &7.123 &6.671 & 0.011 & 2.3 \\
         benzene  & 23.9  & 5.084 & 5.203 & 5.086 & -0.002 & 2.1 \\
         nitrosomethane & 99.5 & 5.772 &9.633 &5.757 &0.015 & 0.37  \\
         nitroxyl  &99.9 & 5.282 &8.863 &5.257 & 0.025 & 0.68  \\
         glyoxal  &99.9 & 6.797 &12.179 &6.763 &0.034 & 0.62 \\
    \end{tabular}
    \caption{Comparison of the RR-EOM-CC3 with the exact EOM-CC3 in aug-cc-pVDZ and aug-cc-pVTZ basis sets for electronic excited states with pronounced doubly-excited character. The size of both ground-state ($N_{\mathrm{svd}}$) and excited-state ($N_{\mathrm{SVD}}$) triple-excitation subspaces were set to the number of molecular orbitals of a given molecule ($N_{\mathrm{MO}}$).  All energy values are in eV.  }
    \label{doubly-excited}
    \end{table}
    {\footnotesize $a$ - excitation energies calculated with the exact EOM-CC3 have been taken from article by Loos et. al.\cite{Loos2022}}
    
    {\footnotesize$\%R_1$ - percentage of single excitations, $\%R_2$ - percentage of double excitations, $\Delta \omega$ -  difference between the RR-EOM-CC3 and the exact EOM-CC3 excitation energies, $\%|\Delta \omega /\omega_{R_3}|$ -  absolute value of the ratio of $\Delta \omega$ and triple amplitudes contribution to the excitation energy (in \%)}

Before discussing in detail the numerical results given in Table~\ref{doubly-excited}, it is important to provide a broader context as to how the EOM-CC3 method performs for different types of excited states. As mentioned in the previous section, for states dominated by single excitations, EOM-CC3 is a very accurate method with average errors of around 0.03\,eV. However, for states of doubly-excited character, the mean absolute error of 0.86 eV has been recently reported by Loos et. al.~\cite{loos2019b} This means that somewhat larger errors would be acceptable for these states also in the rank-reduced formalism. For states of intermediate character, the average EOM-CC3 errors are, understandably, somewhere in between these two extremes.

Let us discuss the RR-EOM-CC3 results from Table~\ref{doubly-excited} taking the aug-cc-pVTZ basis set as an example. The results obtained with the smaller aug-cc-pVDZ basis are only slightly different, on average, and they follow the same trends. For example, the mean absolute errors (|$\Delta \omega$|) are 0.018~eV and 0.019~eV in the aug-cc-pVTZ and aug-cc-pVDZ basis sets, respectively. Therefore, the discussion and conclusions largely apply to the aug-cc-pVDZ results as well. The absolute RR-EOM-CC3/aug-cc-pVTZ errors are within 0.002 - 0.034 eV for systems considered in Table~\ref{doubly-excited} in comparison with the exact EOM-CC3. A somewhat larger error than expected is found in the case of acrolein molecule. However, taking into account that the contribution of doubly-excited configurations is substantial for this state of acrolein, this error is still acceptable. In general, the absolute errors are larger by a factor of $2-3$ for states of pure double-excitation character than for remaining states considered in Table~\ref{doubly-excited} and in the previous section. However, turning our attention to the last column, we see that the values of parameter $\%|\Delta \omega / \omega_{R_3}|$ are much smaller than for the singly-excited states. This is especially pronounced for pure double excitations where the calculated ratio is smaller than 1\%. Both observations are easily explained if one notices how much the exact EOM-CC3 and the exact EOM-CCSD excitation energies differ from each other for the presented set of molecules. This difference is between roughly 0.12 eV (benzene) and 5.4 eV (glyoxal). The errors in the RR-EOM-CC3 excitation energies are thus small in comparison to the overall triple amplitudes contributions, especially for the pure double excitations.

The provided data shows that the RR-EOM-CC3 method is able to deal with states with doubly-excited character with the accuracy close to the exact EOM-CC3 method. This conclusion holds despite the initial guess for the triply-excited amplitudes being based on the EOM-CCSD results which may be wrong by several eV.
    
\subsection{Large molecules}
\label{sec:large}

We can safely say that the usefulness of the RR-EOM-CC3 method rests upon its performance in the case of the large molecules. After all, this is where the reduced $N^6$ scaling of the developed method is the most beneficial and gives hope for a significant computational costs reduction. We carried out calculations of the first excitation energy for the L-proline and heptazine molecules in aug-cc-pVTZ basis set in order to see how much benefit we get from the rank-reduced approach. The results are compared to the exact EOM-CC3 calculations available in the literature.

The exact EOM-CC3 excitation energy for L-proline in aug-cc-pVTZ basis set is equal to 5.72 eV as reported by Paul et. al.~\cite{Paul2021} who tested their efficient implementation of EOM-CC3 in the $e^T$~\cite{eT} program.  As described in the article, this calculation was performed on 44 cores, using 700 GB of memory and lasted approximately 6-7 days (summing up walltimes from all calls to "ground state", "prepare for Jacobian" and "right excited state" calculations, provided by the authors in Table 4). The RR-EOM-CC3 calculation for $N_{\mathrm{SVD}}=N_{\mathrm{MO}}$ results in an error of less than 0.01 eV in excitation energy, while being significantly less computationally demanding. It requires only approximately 140 GB of memory and the calculation lasts approximately 3 days on 24 cores.

Considering the heptazine molecule, the RR-EOM-CC3 calculation ($N_{\mathrm{SVD}}=N_{\mathrm{MO}}$) yields excitation energy of 2.710 eV, while the first excitation energy reported by Loos et. al.~\cite{Loos2023} equals 2.708 eV (calculated in the CFOUR~\cite{cfour, cfour2} program). We note that such a small error is rather surprising and may be accidental -- according to the results reported in the previous section, one should not reasonably expect this level of accuracy to be the norm. However, this does not take away from the fact that the result is promising. Regarding the computational costs, direct comparison with Ref.~\onlinecite{Loos2023} is not straightforward, because the spatial symmetry of heptazine molecule was used in the calculations from Ref.~\onlinecite{Loos2023}, while our implementation is limited at present to the $C_1$ point group. It is well-known that the direct-product decomposition approach~\cite{stanton1991,gauss1991,matthews2019} for exploitation of spatial symmetry in many‐body methods used in calculations from Ref.~\onlinecite{Loos2023} enables to reduce the computational cost by a factor of roughly $\approx g^2$, where $g$ is the order of the molecular point group. The heptazine molecule in its ground state is described by the $D_{3h}$ point group ($g=12$) and hence one can expect that the computational cost is reduced by two orders of magnitude in comparison with the $C_1$ point group. Even exploiting the spatial symmetry, calculations from Ref.~\onlinecite{Loos2023} required about $6$ days per excited state on a machine with 2TB of memory~\cite{loosprivate}. This shows that the canonical EOM-CC3 calculations for heptazine within $C_1$ point group would take an unreasonable amount of time and memory requirements would exceed the capabilities of most machines. In comparison, our method used about 261 GB of memory and the calculations lasted approximately 7 days on 24 cores. Clearly, introduction of the Tucker decomposition scheme results in a much faster calculation which requires significantly less computer memory, while retaining the accuracy of the exact EOM-CC3 method.

\section{Conclusions and Outlook}
\label{sec:outlook}

We report a development of a new method termed rank-reduced equation-of-motion coupled cluster triples (RR-EOM-CC3). It is an approximation to the well-known EOM-CC3 method and relies on the Tucker decomposition of the ground-state and excited-state triple-excitation amplitudes in order to reduce the computational cost of the conventional EOM-CC3 routine. Introduction of this decomposition scheme combined with careful factorization of the working equations enables to reduce the formal scaling of the method to the $N^6$ level. Additionally, the accuracy of the Tucker decomposition, and thus of the RR-EOM-CC3 method itself, can be adjusted through two parameters which define the sizes of the triple-excitation subspaces for the ground state ($N_{\mathrm{svd}}$) and the studied excited state ($N_{\mathrm{SVD}}$) of a given molecule. 

The RR-EOM-CC3 method has been tested and compared to its exact counterpart in a series of calculations of vertical excitation energies, moving along the "diagonal" ($N_{\mathrm{svd}}= N_{\mathrm{SVD}}$) in the aforementioned parameters. We begin our discussion with calculations involving the first excited states of several small molecules, which exhibit single-excitation character. The results indicate that the rank-reduced approach is able to maintain the accuracy of the exact EOM-CC3, provided that one chooses a reasonably large triple-excitation subspace size. We note that the optimal size of the subspace depends on the basis set used in calculations. In aug-cc-pVTZ basis set we recommend to set this parameter equal to the number of molecular orbitals of a studied system (excluding the frozen-core orbitals), $N_{\mathrm{SVD}} = N_{\mathrm{MO}}$, while in aug-cc-pVDZ the optimal value of $N_{\mathrm{SVD}}$ is between $1.35 N_{\mathrm{MO}}$ and  $1.4 N_{\mathrm{MO}}$. In calculations involving singly-excited states, the use of the recommended settings leads to an error approximately six times smaller than the inherent error of the canonical EOM-CC3. Next, we focused on states with pronounced doubly-excited character. It is noteworthy that the RR-EOM-CC3 method is able to handle such states similarly as its canonical counterpart, despite the fact that it utilizes the EOM-CCSD-based guess for the triple-excitation excited-state amplitudes. Although the errors are usually somewhat larger for states of significant double-excitation character, they are in most cases still drastically lower than the difference between EOM-CC3 and FCI. Lastly, we showed how the RR-EOM-CC3 method fares in calculations of the first excitation energy of moderate-size/large molecules, L-proline and heptazine, in comparison to the exact EOM-CC3 results available in the literature. The error for L-proline was smaller than 0.01 eV, while the error for heptazine was equal to only 0.002 eV. Equally importantly, the use of the rank-reduced approach leads to significant time and memory savings. For example, for L-proline we encountered a twofold speed-up of the calculations and memory requirements were reduced roughly fivefold. This shows that the RR-EOM-CC3 method is a promising candidate for calculation of excitation energies of large molecules while maintaining high accuracy provided by approximate inclusion of triply-excited amplitudes. 

There are two natural ways to extend the present work. One of them is the application of the rank-reduced formalism to the EOM-CCSDT theory. The extension to the complete EOM-CCSDT method~\cite{kucharski2001} is expected to provide better description of states with double-excitation character, especially for states where single excitations are nearly absent (pure double excitations)~\cite{loos2019b}. It also opens up a window for future inclusion of quadruple excitations, e.g. at the CC4 level of theory~\cite{kallay2004,kallay2005,loos2021b,Loos2022}. Another possibility is adopting the present RR-EOM-CC3 formalism for triplet spin states. With the recent interest in the singlet-triplet inversion phenomenon~\cite{Loos2023, ehrmaier2019, deSilva2019,Ricci2021,Blaskovits2023,sobolewski2023, ricci2022, drwal2023, sanz-rodrigo2021} such an extension is highly desirable, providing straightforward access to the singlet-triplet energy gap. Note that the value of this gap is crucial to the performance of the proposed next-generation organic light-emitting diodes (OLED) based on molecules exhibiting siglet-triplet inversion~\cite{pollice2021, aizawa2022}. An example of such molecule is heptazine and solid performance of the RR-EOM-CC3 method in the case of the first excited singlet state, as demonstrated in this work, makes the extension of the present research to the triplet states especially enticing.  

\begin{acknowledgement}
The work was supported by the National Science Centre, Poland, under research project 2022/47/D/ST4/01834. We gratefully acknowledge Poland's high-performance Infrastructure PLGrid (HPC Centers: ACK Cyfronet AGH, PCSS, CI TASK, WCSS) for providing computer facilities and support within computational grant PLG/2023/016599.
\end{acknowledgement}

\begin{suppinfo}
The following files are available free of charge:
\begin{itemize}
  \item {\tt Supporting\_Information.pdf}: additional data regarding the impact of the triple-excitation subspace sizes on the accuracy of the results and molecular structures in Cartesian coordinates.
\end{itemize}

\end{suppinfo}

\bibliography{rr-eom-cc3-paper}
\end{document}